\documentstyle[12pt,epsfig,epsf]{article} 
\begin{document}
\baselineskip 16pt plus 2pt minus 2pt
\newcommand{\beq}{\begin{equation}}
\newcommand{\eeq}{\end{equation}}
\newcommand{\beqa}{\begin{eqnarray}}
\newcommand{\eeqa}{\end{eqnarray}}
\newcommand{\dfrac}{\displaystyle \frac}
\renewcommand{\thefootnote}{\#\arabic{footnote}}
\newcommand{\ve}{\varepsilon}
\newcommand{\krig}[1]{\stackrel{\circ}{#1}}
\newcommand{\barr}[1]{\not\mathrel #1}
\newcommand{\vs}{\vspace{-0.29cm}}
\begin{titlepage}


\hfill FZJ--IKP(TH)--1998--12


\begin{center}

\vspace{1.5cm}

{\Large  \bf {Strange magnetism in the nucleon}} \\

\vspace{1.2cm}
                              
{\large
Thomas R. Hemmert,$^{\ddag}$\footnote{email: Th.Hemmert@fz-juelich.de}  
Ulf-G. Mei\ss ner,$^{\ddag}$\footnote{email: Ulf-G.Meissner@fz-juelich.de}
Sven Steininger$^{\dag,\ddag}$\footnote{email: 
S.Steininger@fz-juelich.de}\footnote{Work
supported in part by the Graduiertenkolleg ``Die Erforschung
subnuklearer Strukturen der Materie'' at Bonn University and by the DAAD.}
}

\vspace{0.7cm}

$^{\ddag}${\it Forschungszentrum J\"ulich, IKP (Theorie), D--52425 J\"ulich, Germany}

\vspace{0.3cm}

$^{\dag}${\it Department of Physics, Harvard University, Cambridge MA02138, USA}

\vspace{0.9cm}


\end{center}

\vspace{1.0cm}

\begin{center}

 ABSTRACT

\end{center}

\vspace{0.1cm}

\noindent 
Using heavy baryon chiral perturbation theory to one loop, we derive 
an analytic and parameter--free expression for the momentum dependence
of the strange magnetic form factor of the nucleon $G_M^{(s)} (Q^2)$ 
and its corresponding
radius. This should be considered as a lower bound. We also derive a
model--independent relation between the isoscalar magnetic and the strange
magnetic form factors of the nucleon based on chiral symmetry and SU(3)
only. This gives an upper bound on the strange magnetic form factor. 
We use these limites to derive bounds on the strange magnetic moment
of the proton from the recent measurement of $G_M^{(s)} (Q^2=0.1\,{\rm GeV}^2)$ 
by the SAMPLE collaboration. We stress 
the relevance of this result for the on--going and future experimental 
programs at various electron machines.
\vspace{2.0cm}

\medskip\centerline{
}
\vfill

\end{titlepage}

\noindent
{\bf 1.} There has been considerable experimental and theoretical interest
concerning the question: How strange is the nucleon? Despite
tremendous efforts, we have not yet achieved a detailled understanding about
the strength of the various strange operators in the proton. These are
$\bar{s}s$, as extracted from the analysis of the pion--nucleon
$\Sigma$--term, $\bar{s} \gamma_\mu \gamma_5 s$ as measured e.g. in 
deep--inelastic lepton scattering off protons and the vector current 
$\bar{s} \gamma_\mu s$, which is accesible e.g. in parity--violating
electron--nucleon scattering. A dedicated program at Jefferson
Laboratory preceded by experiments at BATES (MIT) and MAMI (Mainz)
is aimed at measuring the form factors related to the strange vector
current. In fact, the SAMPLE collaboration has recently reported the
first measurement of the strange magnetic moment of the proton~\cite{SAMPLE}. 
To be precise, they give the strange magnetic form factor at a small momentum
transfer, $G_M^{(s)} (q^2=-0.1~{\rm GeV}^2) = +0.23\pm 0.37 \pm 0.15
\pm 0.19\,$nuclear magnetons (n.m.). The rather sizeable error bars document
the difficulty of such type of experiment.  On the theoretical side, 
there is as much or even more uncertainty. To document this, let us pick
one particular approach. Jaffe~\cite{bob} deduced rather sizeable
strange vector current matrix elements from the
dispersion--theoretical analysis of the nucleon electromagnetic form
factors, assuming that the isoscalar spectral functions are dominated
at low momentum transfer by the $\omega$ and $\phi$ mesons. This
analysis was updated in~\cite{hmd} with similar results. However, if
one improves the isoscalar spectral function by considering also the
correlated $\pi\rho$--exchange~\cite{mmsvo} or kaon loops~\cite{hm},
the corresponding strange matrix elements can change dramatically.
Also, the spread of the theoretical predictions for the strange
magnetic moment, $-0.8 \le \mu_p^{(s)} \le 0.5$~n.m.
underlines clearly the abovemade statement (for a review, see ref.\cite{mur}). 
As we will demonstrate in
the following, there is, however, one quantity of reference, namely we
can make a parameter--free prediction for the momentum dependence of
the nucleons' strange magnetic (Sachs) form factor based on the chiral symmetry
of QCD solely. In addition, we derive a leading order
model--independent relation between the strange and the isoscalar
magnetic form factors, which allows to give an upper bound on the
momentum dependence of $G_M^{(s)} (Q^2)$. These two different results
can then be combined to extract a range for the strange magnetic
moment of the proton from the SAMPLE measurement of the form factor at
low momentum transfer.

\medskip

\noindent{\bf 2.}
The strangeness vector current of the nucleon is defined as
\begin{equation}\label{svc}
\langle N|\;\bar{s}\;\gamma_\mu\; s\;|N \rangle
= \langle N|\;\bar{q}\;\gamma_\mu\;
(\lambda^0/3-\lambda^8/\sqrt{3}) \; q\;| N \rangle 
                                  = (1/3)J_{\mu}^0- (1/\sqrt{3})J_{\mu}^8 \; ,
\end{equation}
with $q=(u,d,s)$ denoting the triplet of the light quark fields and
$\lambda^0 = I\; (\lambda^a)$ 
the unit (the $a=8$ Gell--Mann) SU(3) matrix.
Assuming conservation of  all vector currents, the corresponding singlet and octet 
vector current for a spin--1/2 baryon can then be written as (from
here on, we mostly consider the nucleon)
\begin{equation}\label{curr}
J_{\mu}^{0,8}
=\bar{u}_N(p^\prime)\left[F_{1}^{(0,8)}(q^2)\gamma_\mu+ F_{2}^{(0,8)}(q^2)
        \frac{i\sigma_{\mu\nu}q^\nu}{2m_N}\right] u_N(p) \; .
\end{equation}
Here, $q_\mu=p^{\prime}_\mu-p_\mu$ corresponds to the four--momentum transfer to the 
nucleon by the external singlet ($v_{\mu}^{(0)}=v_\mu \lambda^0$) and the octect ($
v_{\mu}^{(8)}=v_\mu\lambda^8$) vector source $v_\mu$, respectively.
The strangeness Dirac and Pauli form factors are defined via:
\begin{equation}
F_{1,2}^{(s)}(q^2)= \frac{1}{3}F_{1,2}^{(0)}(q^2)
-\frac{1}{\sqrt{3}}F_{1,2}^{(8)}(q^2) \; ,
\end{equation}
subject to the normalization $F_1^{(s)} (0) = S_B$, with $S_B$ the
strangeness quantum number of the baryon ($S_N = 0$) and
$F_{2}^{(s)} (0) =\kappa_{B}^{(s)}$, with
$\kappa_{B}^{(s)}$ the  (anomalous) strangeness moment. 
In the following we concentrate our analysis on the ``magnetic'' strangeness
form factor $G_{M}^{(s)}(q^2)$, which in analogy to the (electro)magnetic
Sachs form factor is defined as
\begin{eqnarray}
G_{M}^{(s)}(q^2)=F_{1}^{(s)}(q^2)+F_{2}^{(s)}(q^2)~. \label{eq:def1}
\end{eqnarray}
It is this ``strangeness'' form factor for which chiral perturbation theory 
(CHPT) gives the most interesting
predictions. Furthermore, $G_{M}^{(s)}(q^2)$ is also the strangeness
form factor that figures prominently in the recent Bates
measurement~\cite{SAMPLE}.

\medskip

\noindent {\bf 3.}
Heavy baryon chiral perturbation theory (HBCHPT) is a precise tool to 
investigate the low--energy properties of the nucleon. It has, however, 
been argued that due to the appearance of higher order local contact terms with
undetermined coefficents, CHPT can not be used to make any
prediction for the strange magnetic moment or the strange electric
radius without additional, model--dependent assumptions~\cite{mus}.
However, the analysis of the nucleons electromagnetic form factors
in CHPT shows that to one loop the slope of the isovector
Pauli form factor can be predicted in a parameter--free manner, see 
refs.\cite{BZ,GSS,BKKM,BFHM}. Since to the same order the corresponding
isoscalar piece is a constant, one therefore has a parameter--free prediction 
for the radius of the Pauli form factor.  It is thus natural to extend this
calculation to the three flavor case with the appropriate singlet and
octet currents as defined in Eq.(\ref{curr}).

We give here the relevant HBCHPT Lagrangians needed for the calculation. 
The baryon octet is parametrized in the matrix $B$, which has the usual
transformation properties of any matter field under chiral transformation. We
utilize the chiral covariant derivative $D_\mu$,
\begin{eqnarray}
D_\mu B&=&\left(\partial_\mu +\Gamma_\mu -i v_{\mu}^{(0)}\right) B \\
\Gamma_\mu&=&\frac{1}{2}\left[u^\dagger,\partial_\mu
u\right]-\frac{i}{2}u^\dagger\left(
             v_{\mu}^{(i)}+a_{\mu}^{(i)}\right)u-\frac{i}{2}u\left(v_{\mu}^{(i)}-
             a_{\mu}^{(i)}\right)u^\dagger \; , \quad (i=3,8)
\end{eqnarray}
the chiral vielbein $u_\mu$,
\begin{eqnarray}
u_\mu&=&i\;u^\dagger \nabla_\mu U\; u^\dagger \\
\nabla_\mu U&=&\partial_\mu U-i\left(v_{\mu}^{(i)}+a_{\mu}^{(i)}\right)U+i U
\left(
               v_{\mu}^{(i)}-a_{\mu}^{(i)}\right), \quad (i=3,8)
\end{eqnarray}
where the quantities $v_{\mu}^{(x)} (a_{\mu}^{(x)})\;,x=0,3,8$ correspond to
external vector (axial-vector) sources and $U(x) = u^2 (x)$ parametrizes the
octet of Goldstone bosons. The three flavor HBCHPT Lagrangian then reads
(we only give the terms relevant to the  calculations presented here)
\begin{eqnarray}
{\cal L}_{\rm MB}^{(1)}&=& \langle \bar{B}\;i v^\mu 
D_\mu \;B \rangle +D\;\langle \bar{B}\;S^\mu\{u_\mu,B\}_+ \rangle
               +F\;\langle \bar{B}\;S^\mu[u_\mu,B]_- \rangle \\
{\cal L}_{\rm MB}^{(2)}&=& -\frac{i b_{6a}^F}{4
m_N}\;\langle \bar{B}\left[S^\mu,S^\nu\right]
               [f_{+\mu\nu}^{(3)},B] \rangle-\frac{i b_{6a}^D}{4
m_N}\;\langle\bar{B}\left[S^\mu,S^\nu
               \right] \{f_{+\mu\nu}^{(3)},B\} \rangle \nonumber \\
            & & -\frac{i b_{6b}^F}{4 m_N}\;\langle \bar{B}
               \left[S^\mu,S^\nu\right] [f_{\+\mu\nu}^{(8)},B] \rangle
               -\frac{i b_{6b}^F}{4 m_N}
               \;\langle \bar{B}\left[S^\mu,S^\nu\right]
\{f_{+\mu\nu}^{(8)},B\} \rangle \nonumber \\
            & & -\frac{2 i b_{6c}}{4 m_N}\;
             \langle \bar{B}\left[S^\mu,S^\nu\right]
v_{\mu\nu}^{(0)}\; B \rangle
               +\ldots 
\end{eqnarray}
with
\begin{eqnarray}
f_{+\mu\nu}^{(i)}&=&u^\dagger F_{\mu\nu}^{R\;(i)}u+u
F_{\mu\nu}^{L\;(i)}u^\dagger \\
F_{\mu\nu}^{L,R\;(i)}&=&\partial_\mu F_{\nu}^{L,R\;(i)}-\partial_\nu
F_{\mu}^{L,R\;(i)}
                        -i\left[F_{\mu}^{L,R\;(i)},F_{\nu}^{L,R\;(i)}\right] \\
F_{\mu}^{R\;(i)}&=&v_{\mu}^{(i)}+a_{\mu}^{(i)} \quad\quad
                   F_{\mu}^{L\;(i)}\;=\;v_{\mu}^{(i)}-a_{\mu}^{(i)} \\
v_{\mu\nu}^{(0)}&=&\partial_\mu v_{\nu}^{(0)}-\partial_\nu v_{\mu}^{(0)}~,
\end{eqnarray}
where $\langle \ldots \rangle$ denotes the trace in flavor space and $m_N$ the
nucleon mass. Furthermore,
$F \simeq 1/2$ and $D \simeq 3/4$ are the conventional SU(3) axial
coupling constants (in the chiral limit, to be precise\footnote{To the
order we are working, it is sufficient to identify the physical with the
chiral limit values.}). The dimension two terms are accompanied by finite
low--energy constants (LECs), called $b_{6a}^{D,F},  b_{6b}^{D,F}, b_{6c}$.
Their precise meaning will be discussed later. 

\medskip

\noindent {\bf 4.}
To be specifc, we now consider the strange magnetic form factor to one
loop order in CHPT. The strange magnetic moment of the nucleon gets 
renormalized by the kaon cloud, completely 
analogous to the renormalization of the nucleon isovector  magnetic moment 
$\mu_N$ by the pion cloud~\cite{CP,BKKM,BFHM,JLMS,MS}. It can be written
as
\begin{equation}
\mu_{N}^{(s)} = \mu_{p}^{(s)}=\mu_{n}^{(s)} = \frac{1}{3}G_M^{(0)} (0)
+ \frac{1}{3}b_{6b}^D - b_{6b}^F
                + \frac{m_N M_K}{24\pi F_{\pi}^2}\left(5 D^2-6
                D F+9 F^2 \right)~,
\label{eq:ks}
\end{equation}
with $M_K$  the kaon mass
and $F_\pi \equiv  (F_\pi + F_K)/2 \simeq 102\,$MeV 
the average pseudoscalar decay constant. We use
this value because the difference between the pion and the kaon decay
constants only shows up at higher order. 
One finds that to 
${\cal O}(p^3)$ in the chiral calculation the strange magnetic moments
of the proton and the neutron are predicted to be equal and consist of
three distinct contributions.
First, the singlet magnetic moment $G_M^{(0)} (0)$ is parametrized in
terms of the unknown
singlet coupling $b_{6c}$. It cannot be predicted without additional
experimental input as has already been noted in \cite{mus}. The
counterterms $b_{6b}^{D,F}$, however, can be extracted from the anomalous
magnetic moments $\kappa_p,\kappa_n$ of the proton and the neutron. 
Third, there is a strong
renormalization of $\mu_N$ due to the kaon cloud. To ${\cal O}(p^3)$ we find
$\mu_{N}^{(s)\;{\rm K-loops}}=2.0$, which is large and {\it positive}. 
This result is in
agreement with the calculation of ref.\cite{mus}. It is
well--known that such large leading order kaon loop effects generally
are diminished by higher order corrections (unitarization), see e.g.~\cite{mmsvo}.
We also note that
in some models the strange magnetic moment is assumed to be generated
exclusively by the kaon contributions~\cite{xxx}, which is already
ruled out to leading order chiral analysis of Eq.(\ref{eq:ks}).

To obtain the complete  strange magnetic form factors one only has to 
consider the diagrams shown in 
Fig.1. For the loop graph~1a, in case of an incoming nucleon, the only
allowed intermediate states are $K\Lambda$ and $K\Sigma$, i.e. the
pion and the $\eta$ cloud do not contribute to this order. 
\begin{figure}[t]
\centerline{\epsfig{file=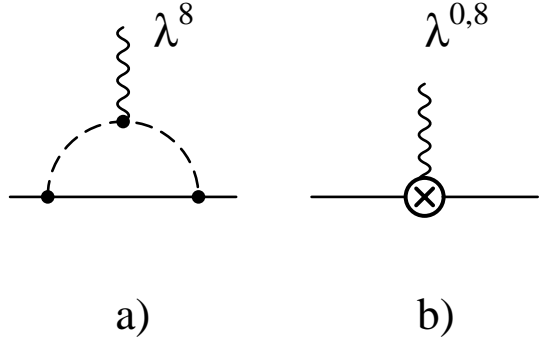,width=2.9in}}

\vspace{0.3cm}

\caption[diag]{\protect \small
Coupling of the singlet ($\sim \lambda^0$)
and octet ($\sim \lambda^8$)
vector currents (wiggly line) to the nucleon
(solid line). a) is a one kaon (dashed line) loop graph and b) a
dimension two contact term. The latter only contributes to the strange
magnetic moment.}
\end{figure}
\noindent For the proton ($p$) and the neutron ($n$) one finds
\begin{eqnarray}\label{eq:c}
G_{M}^{(s)}(Q^2) &=& G_{M}^{(s)\;p}(Q^2)\;=\;G_{M}^{(s)\;n}(Q^2) 
\nonumber \\
                &=& \mu_{N}^{(s)}+\frac{\pi m_N M_K}{(4\pi
F_{\pi})^2}\;\frac{2}{3}\left( 5 D^2-6 D F+9 F^2 \right) \, f(Q^2)~,
\label{eq:gms}
\end{eqnarray}
with $Q^2=-q^2$.  The momentum dependence is given entirely in terms of
\begin{equation}\label{f2q}
f(Q^2) = -\frac{1}{2} + \frac{4+Q^2/M_K^2}{4\sqrt{Q^2/M_K^2}} 
\arctan \biggl( \frac{\sqrt{Q^2}}{2M_K} \biggr)~. 
\end{equation}
The function $f(Q^2)$ is shown in Fig.2. For small and moderate $Q^2$,
it rises almost linearly with increasing $Q^2$.
\begin{figure}[t]
\centerline{\epsfig{file=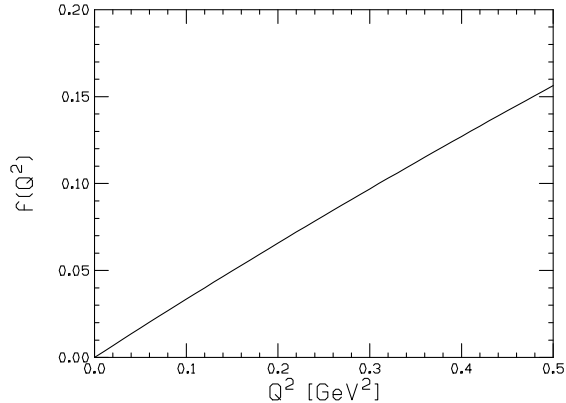,width=2.9in}}

\vspace{0.3cm}

\caption[func]{\protect \small
The function $f(Q^2)$ for small and moderate momentum transfer squared.}
\end{figure}
We note from Eq.(\ref{eq:gms}) that the slope of $G_M^{(s)} (Q^2)$ is
uniquley fixed in terms of well--known low energy parameters,
\begin{equation}
\left(\rho_{M}\right)^2 = \frac{d\; G_{M}^{(s)}(q^2)}{d
  q^2}\biggl|_{q^2=0} =
-\frac{\pi m_N}{(4\pi F_{\pi})^2M_K}\;\frac{1}{18}
\left( 5 D^2-6 D F+9 F^2 \right) 
= -0.027\,\mbox{fm}^2~. \label{eq:rho}
\end{equation}
The slope is identical for a proton or a neutron target, 
it is {\it negative} and to this order {\it independent} of the 
the strange magnetic moment $\mu_{N}^{(s)}$.
The radius has the very reasonable behavior that in the
limit of very heavy kaons, $M_K\rightarrow\infty$, it goes to zero, whereas it
explodes in the chiral limit $M_K\rightarrow 0$. This quantity allows one
to obtain the strange magnetic moment measured at a small value of
$Q^2$ by linear extrapolation to $Q^2 = 0$. Note, however, that the
value given should be considered as a lower limit.  From an analysis of
the electromagnetic form factors of the nucleon we know that at low momentum
transfer the leading CHPT predictions are already quite satisfactory. 
However, in the SU(2) ``small scale expansion framework'' \cite{hhk} it was 
found \cite{BFHM} that the radius of the isovector
magnetic Sachs form factor $G_{M}^{I=1}(q^2)$ is increased by 15-20\% due to
intermediate $\Delta\pi$ cloud effects. As similar analysis in the SU(3)
``small scale expansion'' framework is in preparation to see whether 
there are similarly
sizeable corrections for the magnetic strangeness form factor due to intermediate
decuplet-octet states. In addition, there are other mechanisms (like
e.g. contributions from vector mesons) not covered at this order.

\medskip

\noindent {\bf 5.}
We can also give an upper limit for the strange magnetic form
factor as the following arguments shows. For that, we consider the
electromagnetic current
\begin{eqnarray}
J_{\mu}^{\rm EM}
&=& \langle N|\frac{2}{3}\bar{{\it u}}\gamma_\mu {\it u}-\frac{1}{3}\bar{{\it d}}
              \gamma_\mu {\it d}-\frac{1}{3}\bar{{\it s}}
              \gamma_\mu{\it s}|N\rangle \nonumber
\\
            &=&\frac{1}{2\sqrt{3}} \langle N|\bar{q}\gamma_\mu\lambda_8 q|N\rangle +
               \frac{1}{2} \langle N|\bar{q}\gamma_\mu\lambda_3 q|N\rangle \nonumber \\
            &=&\frac{1}{2\sqrt{3}}J_{\mu}^8+\frac{1}{2} J_{\mu}^3 \; ,
\end{eqnarray}
where $J_{\mu}^8$ corresponds to the octet current of Eq.(\ref{svc}). The
(conserved) triplet current $J_{\mu}^3$ parameterizes the response of a 
nucleon coupled to an external triplet vector source $v_{\mu}^{(3)}=v_\mu \lambda^3$.
The calculation proceeds as before. We find that while in an SU(3) calculation the
magnetic form factors of the proton and the neutron both have a pion
and a kaon cloud contribution, the pion cloud terms drop out to
leading order for the isoscalar magnetic form factor of the nucleon.
Also, in contrast to the SU(2) calculations~\cite{BKKM,BFHM}, the
leading  one loop SU(3) contribution to $G_M^{I=0} (Q^2)$ picks up a
momentum dependence given again entirely  in terms of the function
$f(Q^2)$, see Eq.(\ref{f2q}),
\begin{eqnarray}
G_{M}^{I=0}(Q^2)^{SU(3)}&=&G_{M}^p(Q^2)+G_{M}^n(Q^2) \nonumber \\
                &=&\mu_s-\frac{M_K m_N\pi}{(4\pi F_\pi)^2}\;\frac{2}{3}\left(5
                   \;D^2-6\;D\;F+9\; F^2\right) \, f(Q^2)~,
\label{eq:gmiso}
\end{eqnarray}
with $\mu_s =0.88\,$n.m. the isoscalar nucleon magnetic moment. 
We note that to this order 
in the chiral expansion the prediction is again free of
counterterms for the momentum dependence. Interestingly,
this means that to ${\cal O}(p^3)$ the isoscalar magnetic
form factor of the nucleon is completely dominated by the kaon cloud,
as all  virtual pion
contributions cancel exactly to this order. The result is of course
consistent with the SU(2) analyses of \cite{BKKM,BFHM} as one can check that
$G_{M}^{I=0}(Q^2) \rightarrow \mu_s$ 
in the limit $M_K\rightarrow\infty$, i.e.
the kaon cloud contribution shows up via higher order counterterms in the SU(2)
calculation.
For the leading chiral contribution to the isoscalar magnetic radius one finds
\begin{equation}
\left(r_{M}^{I=0}\right)^2 = \frac{6}{\mu_s}\;\frac{d\;G_{M}^{I=0}(q^2)}
                             {d q^2}\biggl|_{q^2=0} = \frac{\left(5 D^2-6 D F+9
F^2\right) m_N}{48F_{\pi}^2\mu_s  \pi M_K} = 0.18\,\mbox{fm}^2~,
\end{equation}
which is about 27\% of the radius derived from the empirical dipole
parametrization (notice that for the accuracy discussed here, we do
not need to employ more sophisticated parametrizations as e.g. given 
in ref.\cite{MMD}). It should now be clear that the isoscalar
magnetic form factor and the strangeness magnetic form factor of the 
nucleon are closely related. In CHPT one can establish this connection
on a firm ground. Based on the results obtained so far, we can derive 
in addition to the counterterm--free prediction of the low $Q^2$-dependence of
$G_{M}^{(s)}$ in Eq.(\ref{eq:c})  another {\it
model-independent} relation between the isoscalar magnetic form factor 
$G_{M}^{I=0}(q^2)$ of the nucleon and the strange magnetic form factor:
\begin{eqnarray}
G_{M}^{(s)}(Q^2) = \mu_{N}^{(s)}+\mu_s-G_{M}^{I=0}(Q^2)+{\cal O}(p^4)
\label{eq:mi}
\end{eqnarray}
This relation is {\it exact} to ${\cal O}(p^3)$ in SU(3) heavy baryon CHPT. Possible
corrections in
higher orders can be calculated systematically. This relation does not constrain
$G_{M}^{(s)}(0)=\mu_{N}^{(s)}$, but makes new predictions on its
$Q^2$-dependence.
Utilizing the dipole parameterization for $G_{M}^{I=0}(Q^2)$ one finds
\begin{equation}\label{rel}
\left(\rho_{M}^{(s),\;{\rm dip}}\right)^2 = -\frac{d\;
G_{M}^{I=0,\;{\rm dip}}(q^2)}{d q^2}\biggr|_{q^2=0} =  -0.10\, \mbox{fm}^2~.
\end{equation}
This number is roughly three times larger than the leading chiral estimate
of Eq.(\ref{eq:rho}). Given that there are also non--strange contributions
in the isoscalar magnetic form factor, which will start to manifest at
order $q^4$, we consider Eq.(\ref{rel}) as an {\it upper} bound on the strange
magnetic radius.
The corresponding strange magnetic form factor is shown in Fig.3 for
a vanishing strange magnetic moment and using the dipole parametrization
for the isoscalar magnetic form factor. Any finite value for $\mu_N^{(s)}$ can
be accomodated by simply shifting the curve up or down the abzissa.
Note  that a similar dipole--like behaviour with a much smaller slope
(corresponding to the lower bound discussed before) was found in the
vector meson dominance type analysis supplemented by regulated kaon
loops presented in ref.\cite{fork}. It is also important to note that
chiral physics dominates the strange magnetic form factor at low
momentum transfer. However, the steady increase in $G_M^{(s)} (Q^2)$
will eventually be taken over by pole contributions as e.g. exploited
in refs.\cite{bob,hmd} leading to a fall--off at large $Q^2$. At which
momentum transfer that will happen depends on the detailed dynamics
and can only be worked out in specific models, see e.g.~\cite{fork}.
Note that the G0 collaboration will probe this particular range of
momentum transfer~\cite{G0}.
\begin{figure}[t]
\centerline{\epsfig{file=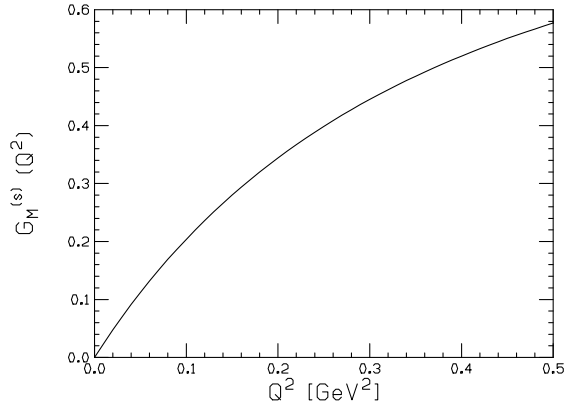,width=2.9in}}

\vspace{0.3cm}

\caption[GMs]{\protect \small
The strange magnetic form factor derived from the isoscalar magnetic one
with $\mu_N^{(s)} =0$.}
\end{figure}

\noindent One can now utilize the $Q^2$--dependence from the two bounds,
Eqs.(\ref{eq:c},\ref{rel}),
to extract the strange magnetic moment from the SAMPLE result for the strange
magnetic form factor. For $Q^2 = 0.1\,$GeV$^2$,
the correction is -0.06 (using $F_\pi = 102\,$MeV) 
and -0.20, respectively, i.e. for the mean value
of ref.\cite{SAMPLE} we get
\begin{equation}
\mu_p^{(s)} = 0.03 \ldots 0.18 \, {\rm n.m.} ~,
\end{equation}
which even for the upper value is a sizeable correction. It is amusing to note
that the small value for $\mu_p^{(s)}$ is in agreement with the analysis presented
in ref.\cite{mmsvo}. Clearly, these numbers should only be considered indicative since
(a) the experimental errors are bigger than the correction and (b) higher order
corrections to the relations derived here should be worked out. 

\medskip

\noindent {\bf 6.}
In summary, we have derived two novel relations which constrain the
momentum dependence of the strange magnetic form factor in the low
energy region. The first one is based on the observation that to one
loop oder in three flavor chiral perturbation theory, the strange form
factor picks up a momentum dependence which is free of unknown
coupling constants, see Eq.(\ref{eq:c}). The second one rests upon the
observation that the isoscalar magnetic form factor calculated in
SU(3) also acquires a momentum dependence which can be related to the
one of the strange magnetic form factor. This gives the
model--independent relation shown in Eq.(\ref{rel}).
These results, which should be considered as a lower and an upper
bound, respectively, should help to sharpen the extraction of
the strange magnetic moment from the measurement of the form factor at
small and moderate momentum transfer. Clearly, the leading order
results discussed here also need to be improved by a systematic
calculation of the corresponding corrections. Such efforts are under way.

  
\bigskip

\noindent {\Large{\bf Acknowledgements}}

\medskip

\noindent 
We would like to thank the participants of the N* workshop at Trento (ECT*) for
helpful comments, especially Steve Pollock.

\bigskip


\end{document}